\documentclass[12pt, a4paper]{iopart}
\usepackage{epsfig, enumerate}
\usepackage{graphicx,epsf}
\usepackage{color}
\usepackage{bm}
\usepackage{iopams}
\eqnobysec
\def\be{\begin{equation}}
\def\ee{\end{equation}}
\def\beb{\begin{equation*}}
\def\eeb{\end{equation*}}
\def\bea{\begin{eqnarray}}
\def\eea{\end{eqnarray}}
\def\beab{\begin{eqnarray*}}
\def\eeab{\end{eqnarray*}}
\def\nn{\nonumber}

\def\p{\partial}

\def\P{{{\cal{P}}}}

\def\H{{\cal H}}

\def\w{{\omega}}
\def\cs2{c_{\rm{s}}^2}

\def \beg {\begin{enumerate}}
\def \en {\end{enumerate}}

\def\Pb{P_0}
\def\rhob{\rho_0}

\def\dPn{{\delta P_{\rm{nad}}}}

\def\cs{c_{\rm{s}}^2}
\newcommand\eq[1]{Eq.~(\ref{#1})}

\def\drho{{\delta\rho_1}}

\def\dP{{\delta P_1}}

\begin{document}

\title{Can Cosmological Perturbations Produce Early Universe Vorticity?}
\author{Adam J.~Christopherson$^*$ and Karim A.~Malik }
\address{Astronomy Unit,
School of Mathematical Sciences, \\
Queen Mary University of London, \\
Mile End Road, London, E1 4NS, United Kingdom.\\
}
\ead{$^*$\mailto{a.christopherson@qmul.ac.uk}}

\begin{abstract}
In this special issue article, based on the talk with the same title in session B5
(Theoretical and Mathematical Cosmology) at GR19, we review the case
of vorticity generation in cosmology using cosmological perturbation
theory. We show that, while at linear order the vorticity evolution equation has no source term
in the absence of anisotropic stress, at second order vorticity is
sourced by gradients in entropy and energy density perturbations. We
then present some estimates for the magnitude and scale dependence of
the vorticity power spectrum using simple input power spectra for the
energy density and entropy perturbations. Finally, we close with
possible directions for future work followed by some hints toward the
observational importance of the vorticity so generated, and the
possibility of primordial magnetic field generation.

\end{abstract}

\maketitle

\section{Introduction}

Vorticity is a common phenomenon in situations involving fluids in the
`real world' (see e.g.~Refs. \cite{landau, Acheson:1990:EFD}). There has also been some interest
recently in studying vorticity in astrophysical scenarios, including
the inter galactic medium \cite{DelSordo:2010mt, Zhu2010}, but
relatively little attention has been paid to the role that vorticity
plays in cosmology and the early universe.\\

In classical fluid dynamics the evolution of an inviscid fluid in the
absence of body forces is governed by the Euler, or momentum, equation
\be
\label{eq:euler}
\frac{\p{\bm v}}{\p t}+({\bm v}\cdot{\bm \nabla}) {\bm v}=-\frac{1}{\rho}{\bm \nabla} P \,,
\ee
where ${\bm v}$ is the velocity vector, $\rho$ the energy density and $P$ the pressure of the fluid.
The vorticity, ${\bm \w}$, is a vector field and is defined as
\be
{\bm \w} \equiv {\bm \nabla}\times{\bm v}\,,
\ee
and can be thought of as the circulation per unit area at a point in the fluid flow. An evolution 
equation for the vorticity can be obtained by taking the curl of \eq{eq:euler}, which gives
\be
\label{eq:vorclassical}
\frac{\p{\bm \omega}}{\p t}
={\bm \nabla}\times({\bm v}\times{\bm \omega})
+\frac{1}{\rho^2}{\bm\nabla}\rho\times{\bm \nabla}P\,.
\ee
The second term on the right hand side of \eq{eq:vorclassical}, often called the baroclinic term in the literature,
then acts as a source for the vorticity. Evidently, this term vanishes if lines of constant energy
and pressure are parallel, or if the energy density or pressure are constant. 
A special class of fluid for which the former is true is a barotropic fluid, defined such that
the equation of state is a function of the energy density only, i.e. $P\equiv P(\rho)$, and so
$(1/ \rho^2){\bm\nabla}\rho\times{\bm \nabla}P=0$.

For a barotropic fluid, the vorticity evolution equation, Eq.~(\ref{eq:vorclassical}), 
can then be written, by using vector calculus identities, as
\be
\label{eq:vorclass2}
\frac{D {\bm \omega}}{D t}\equiv\frac{\p {\bm \omega}}{\p t}+({\bm v}\cdot{\bm \nabla}){\bm \omega}
=({\bm \omega}\cdot{\bm \nabla}){\bm v}-{\bm \omega}({\bm \nabla}\cdot{\bm v})\,,
\ee
which makes it clear that, in this case,
the vorticity vector has no source, and so ${\bm \w={\bm 0}}$ is a solution to 
\eq{eq:vorclass2}. 

Allowing for a more general perfect fluid
with an equation of state depending not only on the energy density, but of the form
 $P\equiv P(S,\rho)$ will mean that, in general, the baroclinic term is no longer vanishing,
and so acts as a source for the evolution of the vorticity. This is Crocco's theorem \cite{crocco} which states that
vorticity generation is sourced by gradients in entropy in classical fluid dynamics.

\section{Elements of Cosmological Perturbation Theory}

In order to study vorticity in the early universe, we must use general relativity. Since Einstein's equations
are notoriously difficult to solve, and relatively few exact, inhomogeneous solutions exist, we 
invoke the technique of cosmological perturbation theory. This means starting with a homogeneous 
and isotropic Friedmann-Robertson-Walker (FRW) spacetime
as a background and adding on top layers of inhomogeneous perturbations. Of course,
perturbing the geometry of the spacetime invokes perturbations in its matter content, so, for example,
the energy density is perturbed as
\be 
\rho(x^i, \eta)=\rhob(\eta)+\delta\rho(x^i, \eta)\,,
\ee
where $\eta$ is conformal time and a subscript zero denotes the homogeneous background part of the 
quantity. The perturbations are then expanded order by order in a series as
\be 
\label{eq:expand}
\delta\rho=\delta\rho_1+\frac{1}{2}\delta\rho_2+\cdots\,,
\ee 
and $\rhob\gg\delta\rho_1>\delta\rho_2$, where the subscripts denote the order of the perturbation.
In analogy with the classical case, we want to consider perturbations of a perfect fluid with equation of 
state  $P\equiv P(S,\rho)$. Expanding the pressure perturbation in a Taylor series gives
\be 
\delta P=\frac{\p P}{\p S}\delta S+\frac{\p P}{\p \rho}\delta\rho\,,
\ee 
which can also be written as 
\be 
\delta P=\dPn+\cs \delta\rho\,,
\ee
where we have introduced the non-adiabatic pressure perturbation, $\dPn$ and the adiabatic sound
speed squared, $\cs$. This can then be extended beyond linear order; see Ref.~\cite{nonad} for details.

The most general perturbations to the FRW line element
are given by \cite{Bardeen:1980kt, ks}
\be
ds^2=a^2(\eta)\left[-(1+2\phi)d\eta^2+2B_idx^id\eta+(\delta_{ij}+2C_{ij})dx^idx^j\right]\,,
\ee
where the perturbations of the spatial components of the metric can be further decomposed 
into scalars, vectors and tensors (characterised by their transformation behaviour on spatial hypersurfaces) 
as
\be
B_i = B_{,i}-S_i \,, \hspace{5mm}
C_{ij} = -\psi \delta_{ij}+E_{,ij}+F_{(i,j)}+\frac{1}{2}h_{ij}\,,
\ee
where $S_i$ and $F_i$ are divergence-free vectors, and $h_{ij}$ is a divergence-free, traceless rank
two tensor. Each metric perturbation is then expanded order by order in an analogous way to the energy density,
Eq.~(\ref{eq:expand}).

Splitting the spacetime into a background and perturbation introduces spurious coordinate artefacts, or 
gauge modes, which have the potential to cause a problem.
In order to remove these gauge modes, we need to make a gauge choice. 
 Since there are myriad articles in the literature dedicated to this issue, we do not further comment
on this here, pointing the interested reader to, for example, 
Refs.~\cite{Bardeen:1980kt, ks, mfb, MW2008} for more information. 
For the rest of this article, we will work in the uniform curvature gauge, for which $E=0=\psi$
and $F_i={\bm 0}$, and neglect tensor perturbations, since they are negligible compared to scalar
perturbations. This gives the line element
\be
ds^2=a^2(\eta)\left[-(1+2\phi)d\eta^2+2B_idx^id\eta+\delta_{ij}dx^idx^j\right]\,,
\ee
and the fluid four-velocity, 
\be
\label{eq:fourvel1}
u_\mu=-\frac{1}{2}a\left[2(1+\phi_1)+\phi_2-\phi_1^2+v_{1k}v_1{}^k,-2V_{1i}-V_{2i}+2\phi_1B_{1i}\right]\,,
\ee
where where $v^i$ is the spatial three-velocity and $V^i\equiv v^i+B^i$.

Evolution and constraint equations are obtained, in general relativity, through the covariant conservation
of energy-momentum, $\nabla_\mu T^\mu{}_\nu=0$, and the Einstein field equations, $G^\mu{}_\nu=8\pi G T^\mu{}_\nu$,
where $T_{\mu\nu}$ is the energy-momentum tensor, and $G_{\mu \nu}$ the Einstein tensor, as usual.
In cosmological perturbation theory, we obtain such equations order by order. As an example, consider
the momentum conservation equation, from $\nabla_\mu T^\mu{}_i=0$, at linear order
\be 
\label{eq:mmtm}
V_{1i}'+\H(1-3\cs)V_{1i}+\left[\frac{\delta P_{1}}{\rhob+\Pb}+\phi_1\right]_{,i}=0\,,
\ee
where $\H$ is the Hubble parameter in conformal time, defined as $\H=a' / a$, the prime denoting a derivative with 
respect to conformal time.
Eq.~ (\ref{eq:mmtm}) is the analogue of the Euler equation (\ref{eq:euler}) in cosmological perturbation theory.

\section{Vorticity in Cosmology}

In general relativity, the vorticity tensor is defined as the
projected anti-symmetrised covariant derivative of the fluid four
velocity, that is \cite{ks}
\be
\label{eq:defomega}
\omega_{\mu\nu}= \P^{~\alpha}_{\mu}
\P^{~\beta}_{\nu} u_{[\alpha; \beta]}
\,,
\ee
where $\P_{\mu\nu}$ is the projection tensor  into the instantaneous fluid 
rest space and is given by
\be
\label{eq:projection}
\P_{\mu\nu}=g_{\mu\nu}+u_{\mu}u_{\nu}.
\ee
Note that, in analogy with the classical case, it is possible to define
a vorticity vector as $\w_\mu=\frac{1}{2}\varepsilon_{\mu\nu\gamma}\w^{\nu\gamma}$,
where $ \varepsilon_{\mu\nu\gamma}$ is the covariant permutation tensor in the fluid rest space
(see Ref.~\cite{Lu2009, vortest}). 

The vorticity tensor can then be decomposed in the usual way,
up to second
order in perturbation theory, as 
$\omega_{ij}\equiv\omega_{1ij} +\frac{1}{2}\omega_{2ij}$. 
Working in the uniform curvature gauge, and considering only
scalar and vector perturbations, we can obtain the components
of the vorticity tensor by substituting the expressions for the 
fluid four velocity, \eq{eq:fourvel1}, along with the metric tensor
 into \eq{eq:defomega}. At first order this gives us
\be
\label{eq:omegafirst}
\omega_{1ij}=aV_{1[i,j]}\,,
\ee
and at second order 
\be
\label{eq:omegasecond}
\w_{2ij}=aV_{2[i,j]}+2a\left[V_{1[i}^\prime V_{1j]}
+\phi_{1,[i}\left(V_{1}+B_{1}\right)_{j]}
-\phi_1B_{1[i,j]}\right] \,.
\ee

\subsection{Vorticity Evolution}

In order to obtain an evolution equation for the vorticity at first order, we take the time derivative of
\eq{eq:omegafirst}. Simplifying, and using the first order equations, gives
\be
V_{1[i,j]}'=-\H(1-3\cs)V_{1[i,j]}=-\frac{1}{a}\H(1-3\cs)\w_{1ij}\,,
\ee
and so, from  \eq{eq:omegafirst}, 
\be 
\omega_{1ij}'-3\H\cs\w_{1ij}=0\,.
\ee
This reproduces the well known result that, during radiation domination, 
$|\w_{1ij}\w_1^{ij}|\propto a^{-2} $ in the absence of an anisotropic stress term \cite{ks}.\\

At second order things are somewhat more complicated. We now need to take the time derivative of 
\eq{eq:omegasecond}, and use the first order evolution and constraint equations, as well as the
second order conservation equations in order to eliminate all the metric perturbation variables.
We omit the derivation and the governing equations in this review article (see Ref.~\cite{vorticity} 
for details of the derivation and the governing equations), 
but calculations give
\bea
\label{eq:vorsecondevofull}
&\w_{2ij}^\prime-3\H\cs\w_{2ij}
+2\left[
\left(
\frac{\dP+\drho}{\rhob+\Pb}\right)^\prime+V^k_{1,k}-X^k_{1,k}
\right]\w_{1ij}  \\
&
+2\left(V^k_1-X^k_1\right)\w_{1ij,k}-2\left(X^k_{1,j}-V^k_{1,j}\right)\w_{1ik}
+2\left(X^k_{1,i}-V^k_{1,i}\right)\w_{1jk}\nn \\
&=\frac{a}{\rhob+\Pb}\Big\{3\H\left(V_{1i}\dPn_{1,j}-V_{1j}\dPn_{1,i}\right)\nn\\
&
\qquad\qquad\qquad+\frac{1}{\rhob+\Pb}\left(\drho_{,j}\dPn_{1,i}-\drho_{,i}\dPn_{1,j}\right)\Big\}\nn \,, 
\eea
where $X_{1i}$ is given entirely in terms of matter perturbations as 
\be
X_{1i}=\nabla^{-2}\left[\frac{4\pi G a^2}{\H}
\left(\drho_{1,i}-\H(\rhob+\Pb)V_{1i}\right)\right]\,.
\ee

In fact, even assuming zero first order vorticity, that is $\w_{1ij}=0$, the second order vorticity evolves as
\be 
\label{eq:vorsecondevolution}
\w_{2ij}^\prime -3\H\cs\w_{2ij}
=\frac{2a}{\rhob+\Pb}\left\{3\H V_{1[i}\dPn_{1,j]}
+\frac{\drho_{,[j}\dPn_{1,i]}}{\rhob+\Pb}\right\}\,,
\ee
 and so we see that there is a non zero source term for the vorticity at second order in perturbation theory which
 is, in analogy with classical fluid dynamics, made up of gradients in entropy and density perturbations. Note that, in 
 the
 absence of a non-adiabatic pressure perturbation, we recover the result of Ref.~\cite{Lu2009} that there is no
 vorticity generation.

\subsection{First Estimates: Magnitude and Scale Dependence}

In this section we present an estimate of the magnitude and scale dependence of the power spectrum for the 
vorticity, assuming simple power-law input power spectra. The results presented here are calculated in 
detail in Ref.~\cite{vortest}.

Working now in the radiation era, and neglecting the vector perturbations (the first term on the right hand side of 
Eq.~(\ref{eq:vorsecondevolution})), we obtain
\bea
\label{eq:vorev2}
\w_{2ij}^\prime -\H\w_{2ij}
=\frac{9a}{8\rhob^2}{\drho_{,[j}\delta P_{{\rm nad}1,i]}}
\equiv S_{ij}
\,,
\eea
and we can define the power spectrum of the vorticity in the usual way, in Fourier space, as
\be
\label{eq:PSomega}
\left<\w_2^*({\bm k}_1, \eta){\w_2}({\bm k}_2, \eta)\right>
=\frac{2\pi}{k^3}\delta({\bm k}_1-{\bm k}_2){\mathcal{P}}_{\w}(k, \eta)\,.
\ee
We take the following input power spectra
\be
\label{input_power_norm}
\drho({k}, \eta)
=A \left(\frac{k}{k_0}\right)\left(\frac{\eta}{\eta_0}\right)^{-4}\,, \hspace{5mm}
\delta P_{\rm nad1}({k}, \eta)= D\left(\frac{k}{k_0}\right)^2\left(\frac{\eta}{\eta_0}\right)^{-5}\,,
\ee
where the time evolution of the energy density is obtained by solving the first order equations, and the wavenumber
dependence from relating an initial ansatz to {\sc Wmap} data. The non-adiabatic pressure input spectrum comes
from demanding that it decays faster than the energy density, and that its spectrum is bluer than that of the
energy density. The amplitudes can also be related to {\sc Wmap} parameters. These input power spectra
allow us to compute the power spectrum for the vorticity analytically, giving
\bea
{\mathcal{P}}_\w(k, \eta)&=
\frac{81}{256}k_0^5\frac{\eta^2}{4 \pi^4}\ln^2\left(\frac{\eta}{\eta_0}\right)
\Bigg(\frac{\alpha(k_0)}{1-\alpha(k_0)}\Bigg)^2k_0^{-12}\Delta_{\mathcal{R}}^8k_{\rm c}^5\nonumber\\
&\qquad
\times\Bigg[\frac{16}{135}\frac{k_{\rm c}^4}{k_0^4}\left(\frac{k}{k_0}\right)^7
+\frac{12}{245}\frac{k_{\rm c}^2}{k_0^2}\left(\frac{k}{k_0}\right)^9
-\frac{4}{1575}\left(\frac{k}{k_0}\right)^{11}\Bigg]\,,
\eea
where $k_{\rm c}$ denotes the large wavenumber or small scale integration cutoff, and the other
parameters are given in Ref.~\cite{WMAP7}. Substituting in an illustrative value for the cutoff,
$k_{\rm c}=10{\rm Mpc}^{-1}$, and using the values of the {\sc Wmap} parameters from
Ref.~\cite{WMAP7}, taking a conservative value of 10\% of the maximum value for $\alpha$, we obtain the following vorticity power spectrum
\bea
{\mathcal{P}}_\w(k, \eta)=\eta^2\ln^2\left(\frac{\eta}{\eta_0}\right)\Bigg[&0.87\times 10^{-2}
\left(\frac{k}{k_0}\right)^7
+3.73\times 10^{-11}\left(\frac{k}{k_0}\right)^9 \nonumber\\
&-7.71\times 10^{-20}\left(\frac{k}{k_0}\right)^{11}\Bigg]{\rm Mpc^{2}}\,.
\eea

This spectrum has a non-negligible magnitude which depends upon the
small scale cut-off $k_{\rm c}$ and the chosen parameters. As this is a second 
order effect the magnitude is somewhat surprising.  As can be seen, the result has 
a dependence on the wavenumber to the power of at least seven for our choice of non-adiabatic
pressure input spectrum.

\section{Future Directions}

In the above section we have presented estimates for the vorticity
power spectrum based on simple, power-law input power spectra. While this is a
good first approximation, in order to obtain more realistic estimates
of the magnitude of the early universe vorticity, we need to go beyond the simple
ansatz for the non-adiabatic pressure perturbation input spectrum.

One way in which a non-adiabatic pressure perturbation can be
generated is through the relative entropy perturbation between two or
more fluids or scalar fields. For example, the relative entropy or
isocurvature perturbation, at first order, between two fluids denoted
with subscripts $A$ and $B$ is \cite{Malik:2002jb}
\be 
{\mathcal S}_{AB}=3\H\Bigg(\frac{\delta\rho_B}{\rho_{0B}'}-\frac{\delta\rho_A}{\rho_{0A}'}\Bigg)\,.
\ee
In a system consisting of multiple fluids, the non-adiabatic pressure perturbation is 
split as \cite{ks,MW2008}
\be 
\delta P_{\rm nad}=\delta P_{\rm intr}+\delta P_{\rm rel}\,,
\ee
where the first term is the contribution from the intrinsic entropy perturbation
of each fluid, and the second term is due to the relative entropy perturbation between each fluid,
${\mathcal S}_{AB}$, and is defined as
\be 
\delta P_{\rm rel}\equiv\frac{1}{6\H \rhob'}
\sum_{A,B}\rho_{0A}'\rho_{0B}'\Big(c_B^2-c_A^2\Big){\mathcal S}_{AB}\,,
\ee
where $c_A^2$ and $c_B^2$ are the adiabatic sound speed of each
fluid. Thus, for a multiple fluid system, even when the intrinsic
entropy perturbation is zero for each fluid, there is a non-vanishing
overall non-adiabatic pressure perturbation. This can be extended to
the case of scalar fields by using standard techniques of treating the
fields as fluids (see e.g.~Ref.~\cite{nonad}). Therefore a
possible next step will involve using the  relative entropy spectrum calculated from
multi-field inflation as an input for the non-adiabatic pressure perturbation. 
This will enable us to obtain a
more realistic description of induced
vorticity in the early universe.

\section{Discussion and Conclusions}

In this brief article, we have reviewed current progress in the generation of vorticity in the 
early universe through non-linear cosmological perturbations. We started out by 
considering the familiar case of classical fluid dynamics and showed that the evolution equation for the classical 
vorticity contains a source term which is only non-zero if the equation of state is a function
of two variables: the energy density and the entropy. We then briefly introduced some elements of
cosmological perturbation theory and, working in the uniform curvature gauge, derived evolution
equations for the vorticity. At linear order we reproduce the well known result that vorticity decays 
with the expansion of the universe in the absence of anisotropic stress. 
However we showed
 that vorticity is sourced is at second order by gradients in energy density 
 and non-adiabatic pressure perturbations. This is in analogy with the classical case. 

 We then presented some first estimates of the power spectrum of the induced vorticity
 by using simple power laws as the input power spectra for the energy density and the entropy
 perturbations. The results show that the magnitude of the vorticity power spectrum
 under this approximation is non-negligible and the amplification due to the large power of 
 $k$ is huge. Finally, we briefly touched on the non-adiabatic pressure perturbation
 created by relative entropy perturbations in multi-fluid systems which will allow us to go
 beyond the simple approximation in the future.\\
 
A non-zero vorticity at second order in perturbation theory has important consequences 
for the generation of magnetic fields, as it has been long known that vorticity and magnetic fields are 
closely related (see Refs.~\cite{biermann, harrison}).
Previous works either used momentum exchange between
multiple fluids to generate vorticity, as in Refs.~\cite{Matarrese:2004kq, Gopal:2004ut, Takahashi:2005nd, 
Ichiki:2006cd, Siegel:2006px, Kobayashi:2007wd, Maeda:2008dv}, 
or used intermediate steps to first generate vorticity for example by using shock fronts 
as in Ref.~\cite{Ryu:2008hi}. However, we do not require such additional steps. 
Therefore, an important extension to the work presented in this review is to
consider the magnetic fields generated by our mechanism which could be an important
step in answering the unknown question regarding the origin of the primordial
magnetic field.

\ack 
AJC is grateful to the organisers of the GR19 conference and  in
particular to the chairs of session B5 for the opportunity to present
this work. The authors thank David Matravers for an enjoyable
collaboration on the work this article is based on. AJC is funded by
a  studentship from the Science and Technology Facilities Council (STFC),
and attendance at the GR19 conference was
supported by the Royal Astronomical Society, the Institute of Physics
and the GR19 Local Organising Committee. KAM is supported, in part,
by STFC under Grant ST/G002150/1.

\section*{References}

\bibliography{ajc}

\bibliographystyle{iopart-num.bst}

\end{document}